\documentclass[conference]{IEEEtran}
\IEEEoverridecommandlockouts
\usepackage[utf8]{inputenc}
\usepackage{bm}
\usepackage{amsfonts}
\usepackage{amssymb}
\usepackage{amsmath}
\usepackage{algorithm}
\usepackage{algpseudocode}
\usepackage{graphicx}
\usepackage{siunitx}
\usepackage{cite}
\usepackage[font=small]{caption}
\usepackage{multicol}

\usepackage[hidelinks]{hyperref}

\title{EMF-Aware Power Control for Massive MIMO: Cell-Free versus Cellular Networks
\thanks{This paper was carried out in the framework of the Horizon Europe project CENTRIC (Grant No. 101096379). The work of Stefano Buzzi was also supported by the EU under the Italian National Recovery and Resilience Plan (NRRP) of NextGenerationEU, partnership on “Telecommunications of the Future” (PE00000001 - program “RESTART”), structural project 6GWINET.}}

\author{\IEEEauthorblockN{Sergi Liesegang$^*$ and Stefano Buzzi$^*\dag$}
\IEEEauthorblockA{$^*$Consorzio Nazionale Interuniversitario per le Telecomunicazioni (CNIT), Parma, Italy \\
$\dag$ University of Cassino, Cassino, Italy  and Politecnico di Milano, Milano, Italy}
E-mails: sergi.liesegang@cnit.it, buzzi@unicas.it}

\begin{document}

\maketitle

\begin{abstract}
The impressive growth of wireless data networks has recently led to increased attention to the issue of electromagnetic pollution. Specific absorption rates and incident power densities have become popular indicators for measuring electromagnetic field (EMF) exposure. This paper tackles the problem of power control in user-centric cell-free massive multiple-input-multiple-output (CF-mMIMO) systems under EMF constraints. Specifically, the power allocation maximizing the minimum data rate across users is derived for both the uplink and the downlink under EMF constraints. The developed solution is also applied to a cellular mMIMO system and compared to other benchmark strategies. Simulation results prove that EMF safety restrictions can be easily met without jeopardizing the minimum data rate, that the CF-mMIMO outperforms the multi-cell massive MIMO deployment, and that the proposed power control strategy greatly improves the system fairness.
\end{abstract}


\section{Introduction} \label{sec:1}
Given the impressive growth of wireless communications, health risks related to the electromagnetic field (EMF) are becoming a growing concern worldwide \cite{Zap22}. Indeed, to comply with the increasing demand for broadband services, denser deployments, and higher frequencies are mandatory \cite{Tat21}, which calls for the need to impose EMF exposure constraints in the design of wireless networks \cite{Cas21}. 

According to the most recent studies reported by the International Commission on Non-Ionizing Radiation Protection (ICNIRP) and the United States Federal Communications Commission (FCC), several criteria can be established to limit user radiation to avoid any alleged medical condition (cf. \cite{Chi21}). This radiation is generally modeled through (i) the specific absorption rate (SAR) but also (ii) the incident power density (IPD) or maximum permissible exposure \cite{ITU20}. All these metrics capture the characteristics of the propagation environment and measure the EMF radiation perceived (i) by the human body (or certain parts) in W/kg and (ii) over a specific coverage area in W/m\textsuperscript{2}. For short distances (less than 20 cm) and low frequencies (typically below 6 GHz), deeper skin penetrations are experienced. That is why SAR usually dominates the radiation exposure in the uplink (UL) \cite{Pso22}. Contrarily, IPD restrictions are commonly applied in downlink (DL) transmissions, where EMF absorption is mostly superficial \cite{ICNIRP20}.

One of the key 5G wireless technologies has been the usage of base stations (BSs) equipped with a large number of transmitting antennas, generally referred to as \textit{massive multiple-input-multiple-output} (mMIMO) \cite{Lu14}. In such systems, edge users suffer from inter-cell interference and poor channel conditions that rapidly degrade performance \cite{Buz17}. Due to such attenuation, the DL contribution to EMF exposure is often omitted \cite{Ibr22}. To alleviate the poor performance experienced by cell-edge users, it has been proposed to deploy the antennas in a distributed manner, i.e., through access points (APs), and to design the network in a user-centric fashion \cite{Int19}. This new paradigm, entitled \textit{cell-free} mMIMO (CF-mMIMO), eliminates cell borders and guarantees good quality of service (QoS) for all users \cite{Wia23}. CF-mMIMO is one of the prominent technologies for future 6G data networks.

In CF-mMIMO, however, DL radiation might no longer be negligible since APs are closer to users (the coverage area is reduced), and the resulting path loss might be small. At the same time, however, less power is needed in the UL (users are served by various APs), and, thus, the perceived exposure can be potentially reduced as well. This trade-off makes the performance of CF-mMIMO under safety constraints unclear and naturally raises the comparison with previous centralized scenarios. To the best of the authors' knowledge, this analysis is still unexplored and will be covered in this work, where we will configure the power control in the DL and UL to satisfy both QoS and EMF requirements. 

This paper is structured as follows. Section~\ref{sec:2} introduces the system models and the EMF exposures for DL and UL. Section~\ref{sec:3} formulates the QoS optimization problems, while their solution is presented in Section~\ref{sec:4}. Section~\ref{sec:5} is devoted to the numerical experiments. Section~\ref{sec:6} concludes the work.

\section{System Model} \label{sec:2}
\begin{figure}[t]
    \centerline{\includegraphics[trim={0 0 1cm 1.25cm},clip=true, scale = 0.25]{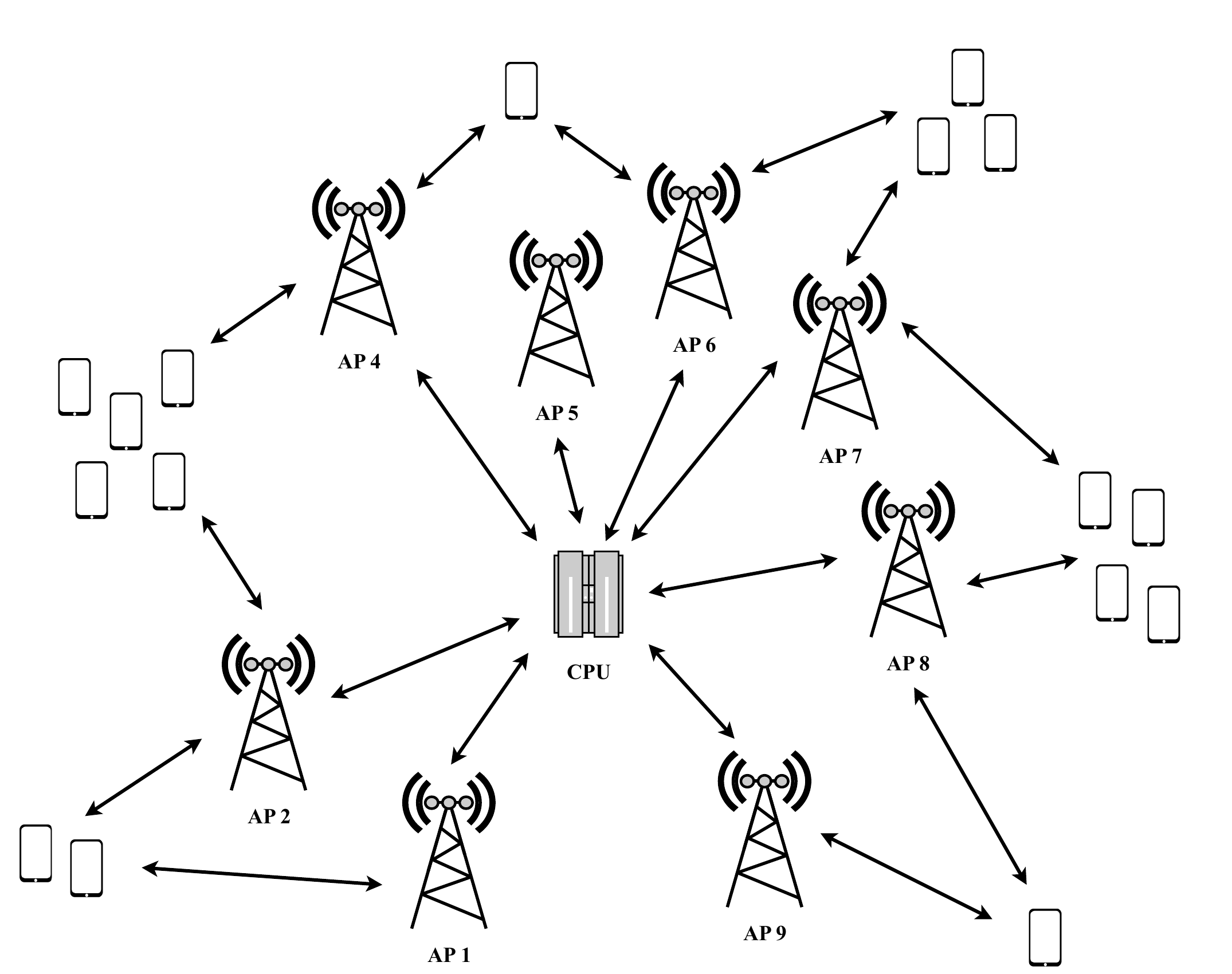}}
    \caption{Illustrative example of a user-centric CF-mMIMO setup with $K = 14$ users and $M = 9$ APs equipped with $3$ antennas. In this case, users are connected to only $N = 2$ APs.}
    \label{fig:1}
    \vspace{-5mm}
\end{figure}

We consider a CF deployment with a group of $K$ single-antenna users served by a set of $M$ APs. All APs have $L$ antennas and are connected to a central processing unit (CPU) through fronthaul links with unlimited capacity\footnote{Since we focus on the impact of radiation constraints, the corresponding extension to a capacity-limited fronthaul is out of the scope of this study.}. An example is depicted in Fig.~\ref{fig:1}. As discussed below, only a subset of $N$ APs might be associated with each user. 

\subsection{Downlink Transmission}
Under a non-orthogonal communication, the discrete-time signal  transmitted from the $m$-th AP can be written as \cite{Dem21}%
\begin{equation}
    \mathbf{x}_m = \sum_{k = 1}^K a_{k,m} \sqrt{p_{k,m}} \mathbf{b}_{k,m}s_k,
    \label{eq:1}
\end{equation}
where $a_{k,m} \in \{0,1\}$ represents the association between user $k$ and AP $m$, i.e., $a_{k,m} = 1$ whenever they are connected and $a_{k,m} = 0$ otherwise. In line with the frequently adopted scheme from the literature, only the $N$ links with the highest large-scale fading coefficients are activated \cite{Elw23}. Additionally, $\mathbf{b}_{k,m} \in \mathbb{C}^{L}$ is the unit-norm beamforming vector\footnote{As discussed in Section~\ref{sec:5}, we concentrate on typical schemes and leave the optimal processing for future studies.} used by the $m$-th AP to direct the information signal $s_k$ (shared among all the APs) towards the $k$-th user. We also assume that $s_k$ follows a complex Gaussian distribution with zero mean and unit power, i.e., $s_k \sim \mathcal{CN}(0,1)$. Finally, $p_{k,m}$ denotes the DL power control coefficient (which will be designed later on).

The received signal at the $k$-th user is given by%
\begin{equation}
    y_k = \sum_{m = 1}^M \mathbf{h}_{k,m}^{\textrm{H}}\mathbf{x}_m + w_k,
    \label{eq:2}
\end{equation}
where $\mathbf{h}_{k,m} \in \mathbb{C}^{L}$ is the UL channel\footnote{Like in most CF-mMIMO systems, channel reciprocity can be exploited thanks to the use of time-division duplexing (cf. \cite{Buz17}).} from user $k$ to AP $m$ and $w_k$ is the complex additive white Gaussian noise with variance $\sigma_k^2$, i.e., $w_k \sim \mathcal{CN}(0,\sigma_k^2)$. Based on the above notation, and assuming independent transmit signals, the signal-to-interference-plus-noise ratio (SINR) at the intended user $k$ can be shown to be written as\footnote{This (instantaneous) SINR expression is valid as long as perfect (local) channel state information (CSI) is considered \cite{Int19}. Since the main goal of the present work is to provide a preliminary outlook of CF-mMIMO performance under EMF constraints, we first concentrate on this assumption for simplicity and later introduce estimation errors in the numerical simulations.}%
\begin{equation}
    \gamma_k = \frac{\displaystyle \left\vert \sum_{m = 1}^M a_{k,m} \sqrt{p_{k,m}} \mathbf{h}_{k,m}^{\textrm{H}} \mathbf{b}_{k,m} \right\vert^2}{\displaystyle \sum_{j \neq k}\left\vert \sum_{m = 1}^M a_{j,m} \sqrt{p_{j,m}} \mathbf{h}_{k,m}^{\textrm{H}} \mathbf{b}_{j,m} \right\vert^2 + \sigma_k^2}.
    \label{eq:3}
\end{equation}

Given the SINR expression in \eqref{eq:3}, we can also model the EMF user exposure in the DL. In particular, omitting the negligible power coming from the noise, the IPD at user $k$'s location includes the contribution of the desired and interfering signals \cite{Pso22, Gon23}:%
\begin{equation}
    \xi_k = \frac{4 \pi}{\lambda^2} \sum_{j = 1}^K \left\vert \sum_{m = 1}^M a_{j,m} \sqrt{p_{j,m}} \mathbf{h}_{k,m}^{\textrm{H}} \mathbf{b}_{j,m} \right\vert^2,
    \label{eq:4}
\end{equation}
where $\lambda = c/\kappa$ is the carrier wavelength, with $\kappa$ the system's operating frequency and $c$ the speed of light.

\subsection{Uplink Transmission}
The received signal at the $m$-th AP is \cite{Elw23}%
\begin{equation}
    \mathbf{r}_m = \sum_{k = 1}^K \mathbf{h}_{k,m}z_k + \mathbf{n}_m,
    \label{eq:5}
\end{equation}
where $z_k$ is the signal transmitted by the $k$-th user, normally distributed with zero mean and power $q_k$, i.e., $z_k \sim \mathcal{CN}(0,q_k)$, and $\mathbf{n}_m$ is the thermal noise, i.e., $\mathbf{n}_m \sim \mathcal{CN}(\mathbf{0}_{L},\eta_m^2 \mathbf{I}_{L})$.

At the detection stage, the $m$-th AP first applies a (linear) spatial filter $\mathbf{f}_{k,m} \in \mathbb{C}^{L}$ to locally decode the message that is coming from the $k$-th user. Next, each AP sends the processed information to the CPU for obtaining the final estimate, i.e., $\hat{z}_k = \sum_{m=1}^M a_{k,m} \mathbf{f}_{k,m}^{\textrm{H}} \mathbf{r}_m$. Hence, the SINR is directly \cite{Dem21}%
\begin{equation}
    \rho_k = \frac{\displaystyle q_k \left\vert \sum_{m = 1}^M a_{k,m} \mathbf{f}_{k,m}^{\textrm{H}}  \mathbf{h}_{k,m} \right\vert^2}{\displaystyle \sum_{j \neq k}q_j\left\vert \sum_{m = 1}^M a_{k,m} \mathbf{f}_{k,m}^{\textrm{H}} \mathbf{h}_{j,m} \right\vert^2 + \sum_{m = 1}^M a_{k,m} \eta_m^2 \left\| \mathbf{f}_{k,m} \right\|_2^2 }.
    \label{eq:6}
\end{equation}

Finally, as already mentioned before, the UL radiation can be modeled via different SARs \cite{Zap22, Cas21}, namely%
\begin{equation}
    \varepsilon_{k,n} = b_{k,n} q_k,
    \label{eq:7}
\end{equation}
where $b_{k,n}$ is the coefficient associated to the $n$-th body part of user $k$ (e.g., head, chest, etc.). 

\section{Problem Formulation}\label{sec:3}
This work aims to design the power control to maximize the minimum data rate in the DL and UL (equivalent to ensuring a certain QoS for all users) under EMF constraints. In both cases, the optimization problem can be formulated as%
\begin{equation}
    \underset{{\mathcal{P}}}{\textrm{max}} \, \underset{k}{\textrm{min}} \, \, R_k\left(\mathcal{P}\right) \quad \textrm{s.t.} \quad \mathcal{C}\left(\mathcal{P}\right),
    \label{eq:8}
\end{equation}
where $\mathcal{P}$ refers to the transmit powers, $R_k(\mathcal{P})$ is the data rate, and $\mathcal{C}(\mathcal{P})$ denotes the constraint set. 

\subsection{Downlink Scenario}
In this scenario, $\mathcal{P} = \{p_{k,m}\}$ contains the APs' transmit powers. Accordingly, the user data rate reads as follows:%
\begin{equation}
    R_k\left(\mathcal{P}\right) = \frac{\tau_d}{\tau_c} B \log_2 \left(1 + \gamma_k\left(\{p_{k,m}\}\right)\right),
    \label{eq:9}
\end{equation}
and the constraint set $\mathcal{C}(\mathcal{P})$ can be represented by%
\begin{equation}
    \begin{aligned}
        C1:& \quad p_{k,m} \geq 0 \quad \forall k,m \\
        C2:& \quad \sum_{k = 1}^K p_{k,m} \leq P_m \quad \forall m \\
        C3:& \quad \xi_k \leq I_k \quad \forall k. 
    \end{aligned}
    \label{eq:10}
\end{equation}

Note that, to calculate the rate in \eqref{eq:9}, we assumed a common bandwidth $B$ and a coherence time $\tau_c$ that splits into $\tau_d$ (DL) and $\tau_u$ (UL) such that $\tau_d + \tau_u \leq \tau_c$ \cite{Elw23}.  

Regarding the constraints in \eqref{eq:10}, $C1$ ensures non-negative coefficients (true by definition), $C2$ introduces the set of power budgets $P_m$ for the different transmitters (i.e., APs), whereas $C3$ limits the radiation perceived at the user side to $I_k$. This will be more carefully discussed later in Section~\ref{sec:5}, where the parameters above will be specified.

\subsection{Uplink Scenario}
Similarly, here we have $\mathcal{P} = \{q_k\}$,%
\begin{equation}
    R_k\left(\mathcal{P}\right) = \frac{\tau_u}{\tau_c} B\log_2 \left(1 + \rho_k\left(\{q_k\}\right)\right),
    \label{eq:11}
\end{equation}
and, thus, $\mathcal{C}(\mathcal{P})$ can be defined with the constraints%
\begin{equation}
    \begin{aligned} 
        C1:& \quad 0 \leq q_k \leq Q_k \quad \forall k \\
        C2:& \quad \varepsilon_{k,n} \leq E_{k,n} \quad \forall k,n,
    \end{aligned}
    \label{eq:12}
\end{equation}
where $C1$ bounds the support of the powers of all the users between $0$ and $Q_k$, while $C2$ fixes the maximum value of the EMF exposure corresponding to the $n$-th body part at $E_{k,n}$.

\section{Proposed Solutions}\label{sec:4}
The convexity of the optimization problem depends on the data transmission, DL or UL. This is addressed in the sequel, where we will distinguish between these configurations.

In the DL, given the nature of \eqref{eq:8}, we must resort to suboptimal approaches for finding a feasible solution. That is why, in the upcoming subsection, we present an iterative procedure that relies on successive convex approximation (SCO) and leads to a stationary point \cite{Lie23}. More specifically, at the $i$-th iteration, the power set $\mathcal{P} \equiv \mathcal{P}^{(i)}$ is updated from the previous solution $\mathcal{P}^{(i -1)}$ until convergence is reached. To do so, we first transform the original formulation by moving the objective function to a new constraint and, after that, approximate the resulting set $\mathcal{C}(\mathcal{P})$ with suitable convex functions. 

Contrarily, as we will see next, the resulting problem is convex in the UL setup. As a result, a globally optimal solution can be found via classic optimization methods \cite{Ngo17}. 

\subsection{Downlink Power Control} \label{sec:4.1}
Let us consider $\phi_{k,m} \triangleq \sqrt{p_{k,m}} \geq 0$ as the new design variable for the DL optimization. Thereby, it can be shown that problem \eqref{eq:8} is equivalent to (cf. \cite{Boy04})%
\begin{equation}
    \begin{aligned}
        \underset{\{\phi_{k,m}\},\delta}{\textrm{max}} \, \, & \delta \\
        \textrm{s.t.} \quad & C1: \phi_{k,m} \geq 0 \quad \forall k,m\\ 
        &C2: \sum_{k = 1}^K a_{k,m} \phi_{k,m}^2 \leq P_m \quad \forall m \\ 
        &C3: \frac{4 \pi}{\lambda^2} \sum_{j = 1}^K \left\vert \sum_{m = 1}^M a_{j,m} \phi_{j,m} \mathbf{h}_{k,m}^{\textrm{H}} \mathbf{b}_{j,m} \right\vert^2 \leq I_k \quad \forall k \\
        &C4: \displaystyle\frac{\tau_d}{\tau_c} B \log_2 \left(1 + \gamma_k\left(\{\phi_{k,m}\}\right) \right) \geq \delta \quad \forall k,
    \end{aligned}
    \label{eq:13}    
\end{equation}
where we introduced an auxiliary variable $\delta$ to transform the non-concave objective function as constraint $C4$, which is still nonconvex. In light of that, we proceed like so.

Since the logarithm is a monotonically increasing function, $C4$ can be reformulated in the following manner:%
\begin{equation}
    \begin{aligned}
        \underbrace{\delta \left(\sum_{j \neq k}\left\vert \sum_{m = 1}^M a_{j,m} \phi_{j,m} \mathbf{h}_{k,m}^{\textrm{H}} \mathbf{b}_{j,m} \right\vert^2 + \sigma_k^2\right)}_{\triangleq f_k\left(\left[\bm{\phi}_1,\ldots,\bm{\phi}_{k - 1},\bm{\phi}_{k + 1},\ldots,\bm{\phi}_K\right]\right)}& \\
        - \underbrace{\left\vert \sum_{m = 1}^M a_{k,m} \phi_{k,m} \mathbf{h}_{k,m}^{\textrm{H}} \mathbf{b}_{k,m} \right\vert^2}_{\triangleq f_k\left(\bm{\phi}_k\right)}& \leq 0 \quad \forall k,
    \end{aligned}
    \label{eq:14}
\end{equation}
which bounds a difference of convex functions with respect to (w.r.t.) the ``amplitude'' vectors $\bm{\phi}_k \triangleq [\phi_{k,1},\ldots,\phi_{k, M}]^{\textrm{T}}$.

Unfortunately, there is no analytic closed-form solution for problems involving this type of nonconvex constraint. With the help of SCO, though, we can find a local optimum. In a nutshell, for a given $\delta$, the optimization will be decomposed into a sequence of subproblems solved iteratively. Each of them needs to have a global optimum for guaranteeing convergence, which means the second term in $C4$ must be approximated by a surrogate function \cite{Sun17}. On top of that, a bisection search can be applied to derive the optimal $\delta$ \cite{Ngo17}.

Among others, a widely employed strategy is to linearize the function $f_k(\bm{\phi}_k)$ so that the constraint in \eqref{eq:14} is convexified. In particular, when applying the first-order Taylor expansions at the previous feasible point, i.e., $\bm{\phi}_k^{(i-1)}$, we can obtain the following lower bound:%
\begin{equation}
    f_k\left(\bm{\phi}_k\right) \geq f_k\left(\bm{\phi}_k^{(i-1)}\right) + \nabla f_k\left(\bm{\phi}_k^{(i-1)}\right)^{\textrm{T}}\left(\bm{\phi}_k- \bm{\phi}_k^{(i-1)}\right),
    \label{eq:15}
\end{equation}
where the gradient $\nabla f_k(\bm{\phi}_k)$ is%
\begin{equation}
    \nabla f_k(\bm{\phi}_k) = 2 \textrm{Re}\left\{\mathbf{g}_k\mathbf{g}_k^{\textrm{H}}\right\}\bm{\phi}_k,
    \label{eq:16}
\end{equation}
with $\mathbf{g}_k \triangleq [a_{k,1}\mathbf{h}_{k,1}^{\textrm{H}} \mathbf{b}_{k,1},\ldots, a_{k, M}\mathbf{h}_{k, M}^{\textrm{H}} \mathbf{b}_{k, M}]^{\textrm{T}}$. 

\pagebreak

Accordingly, at each iteration, we will end up with a subproblem that is a worst-case scenario (more restrictive constraint) but can be globally solved with standard numerical routines, e.g., CVX \cite{CVX20}. Finally, the whole procedure is repeated until we converge to a local optimum.

\subsection{Uplink Power Control} \label{sec:4.2}
Following similar steps, the power control during the UL transmission becomes%
\begin{equation}
\begin{aligned}
    \underset{\{q_k\},\delta}{\textrm{max}} \, \, & \delta \\
    \textrm{s.t.} \quad & C1: 0 \leq q_k \leq Q_k \quad \forall k \\ 
    &C2: b_{k,n} q_k \leq E_{k,n} \quad \forall k,n \\
    &C3: \displaystyle\frac{\tau_u}{\tau_c} B \log_2 \left(1 + \rho_k\left(\{q_k\}\right) \right) \geq \delta \quad \forall k.
\end{aligned}
\label{eq:17}    
\end{equation}

Unlike before, the newly added constraint $C3$ is quasiconvex (lower bound on the logarithm of a linear fraction; thus, quasiconcave). Accordingly, since $C1$ and $C2$ are linear, for every $\delta$, this will result in a quasilinear problem whose (global) optimum can be found. Once again, the optimal value of $\delta$ can be obtained via the bisection method \cite{Boy04}.

\section{Numerical Simulations}\label{sec:5}
Numerical simulations should compare the proposed setting with the following two baseline schemes: (a) a CF-mMIMO system optimized according to problem \eqref{eq:8} without constraint $C3$ (DL) or $C2$ (UL); this permits understanding the impact, on the system's performance, of the EMF constraint; and (b) a multi-cell mMIMO (MC-mMIMO) system where each user is connected to only one macro BS; in this case, we will be able to evaluate the possible advantages of the CF-mMIMO deployment in fulfilling EMF constraints.

\begin{figure}[t]
    \centerline{\includegraphics[trim={0 1.75cm 1cm 1.25cm}, clip=true, scale=0.25]{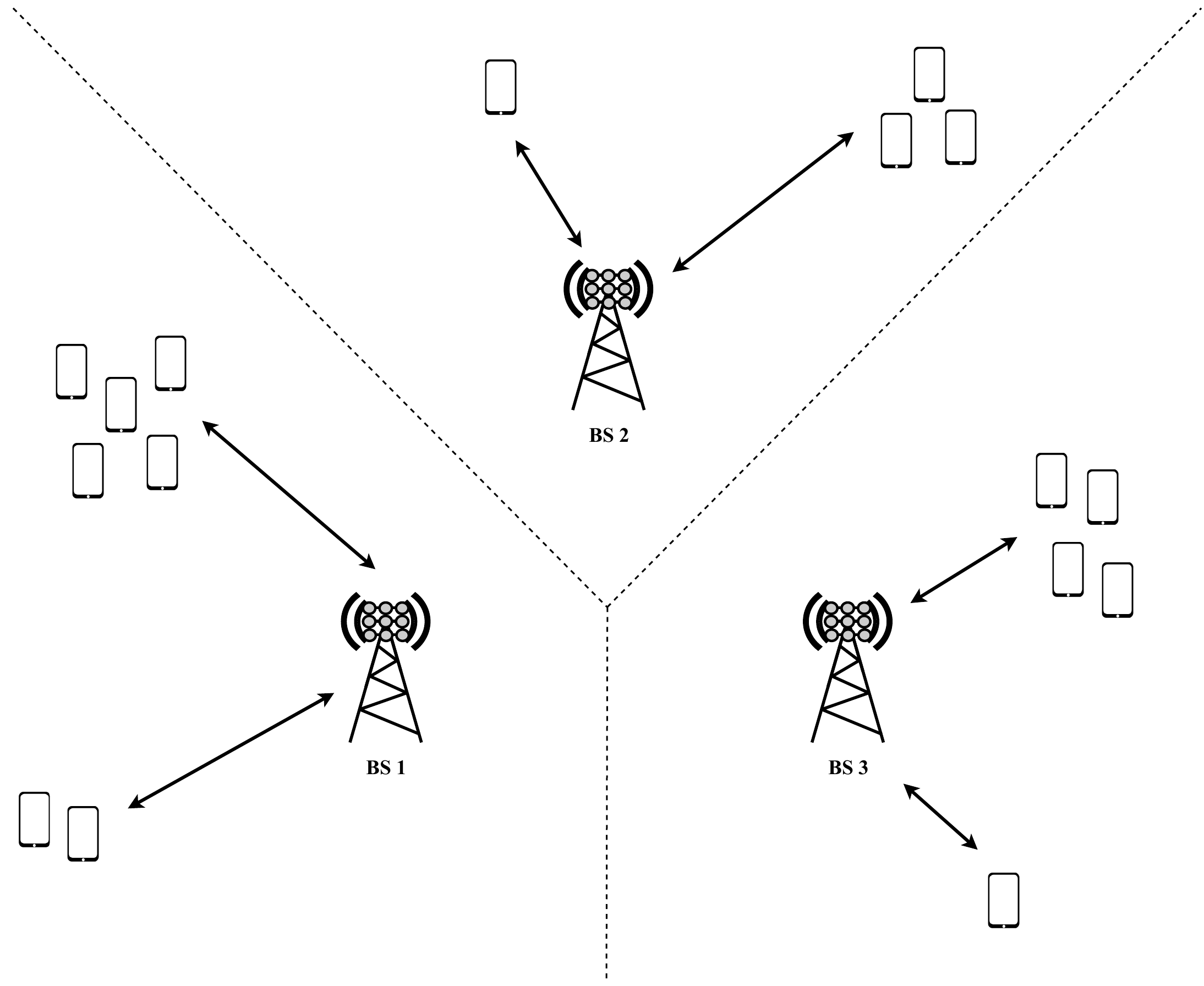}}
    \caption{Illustrative example of a MC-mMIMO setup with $K = 14$ users and $L = 3$ multi-antenna BSs. Unlike cell-free, users are connected to only one BS (dashed lines indicate cell borders).}
    \label{fig:2}
    \vspace{-3mm}
\end{figure}

For fairness, the cellular setup will comprise $L$ BSs with $M$ antennas. This is illustrated in Fig.~\ref{fig:2}, where the numerology is consistent with Fig.~\ref{fig:1}.

\subsection{Propagation Channel}
The link between user $m$ and AP $k$ is \cite[(1)]{Elw23}%
\begin{equation}
    \mathbf{h}_{k,m} = \sqrt{\frac{\alpha_{k,m}}{1 + \beta_{k,m} }}\left(\bar{\mathbf{h}}_{k,m} + \sqrt{\beta_{k,m}}e^{j \psi_{k,m}} \mathbf{v}_m\left(\theta_{k,m}\right) \right),
    \label{eq:18} 
\end{equation}
where $\alpha_{k,m}$ is the large-scale fading coefficient including the path loss, $\beta_{k,m}$ is the Rician factor, $[\bar{\mathbf{h}}_{k,m}]_l \sim \mathcal{CN}(0,1)$ are the uncorrelated Rayleigh distributed non-line-of-sight (NLoS) components, $\psi_{k,m} \sim \mathcal{U}[0,2\pi]$ is the phase offset, $\mathbf{v}_m(\cdot) \in \mathbb{C}^{L}$ is the steering vector (generated according to a uniform linear array), and $\theta_{k,m}$ is the corresponding (LoS) angle of arrival.

Following the discussion in \cite{DAn20}, the Rician factors depend on the probability of LoS of each link, i.e.,%
\begin{equation}
    \beta_{k,m} = \frac{p_{\textrm{LoS}}\left(\zeta_{k,m}\right)}{1 - p_{\textrm{LoS}}\left(\zeta_{k,m}\right)},
    \label{eq:19}
\end{equation}
where $p_{\textrm{LoS}}(\zeta_{k,m})$ is a function of the distance $\zeta_{k,m}$ from user $k$ to AP $m$ \cite[Table B.1.2.1-2]{3GPP36814}.

Regarding the channels between users and macro BSs, we adopt a Rician modeling. However, to avoid redundancy, their derivation is omitted. 

\subsection{Linear Uplink Channel Estimation}
Perfect CSI might be an unrealistic assumption in practical systems. Instead, we must acquire this knowledge locally at the APs through UL orthogonal pilots. This will allow us to characterize the sufficient statistics of the channels \cite{Ngo17}. 

After some manipulations, the following linear minimum mean-squared error estimates can be constructed:%
\begin{equation}
    \hat{\mathbf{h}}_{k,m} = \sqrt{\mu_k}\mathbf{C}_{k,m} \mathbf{D}_{k,m}^{-1} \mathbf{u}_{k,m},
    \label{eq:20}
\end{equation}
where $\mu_k$ is the training pilot's power,%
\begin{equation}
    \mathbf{C}_{k,m} = \frac{\alpha_{k,m}}{1 + \beta_{k,m} } \left(\mathbf{I}_L + \beta_{k,m} \mathbf{v}_m\left(\theta_{k,m}\right)\mathbf{v}_m^{\textrm{H}}\left(\theta_{k,m}\right) \right),
    \label{eq:21}
\end{equation}
refers to the covariance matrix of $\mathbf{h}_{k,m}$, and%
\begin{equation}
    \mathbf{D}_{k,m} = \sum_{j = 1}^K \mu_j \bm{C}_{j,m} \left\vert \mathbf{t}_j^{\textrm{H}} \mathbf{t}_k \right\vert^2 + \eta_m^2 \mathbf{I}_L,
    \label{eq:22}
\end{equation}
is the covariance matrix of the observation $\mathbf{u}_{k,m} \in \mathbb{C}^L$, with $\mathbf{t}_k \in \mathbb{C}^{\tau_p}$ the training sequence of length $\tau_p$ sent by user $k$. Please refer to \cite[Subsection~II-C]{DAn20} for more details.

This way, we can incorporate CSI errors into the QoS power control. In short, we will replace the channels in the SINRs \eqref{eq:3} and \eqref{eq:6} by their estimates and investigate the performance under conjugate beamforming\footnote{Due to space limitations, the use of more advanced processing techniques such as zero-forcing will be the object of study of future works.} (also known as matched filters or maximum ratio), i.e., $\mathbf{b}_{k,m} = \mathbf{f}_{k,m} = \hat{\mathbf{h}}_{k,m}/\| \hat{\mathbf{h}}_{k,m} \|$ \cite{Dem21}.

\subsection{System Parameters}
Throughout all experiments, we consider a deployment area of $1$ km\textsuperscript{2}, wrapped around the edges to avoid boundary effects. The scenario follows the micro-urban configuration described in \cite{3GPP36814} with $P_m = P = 23$ dBm $\forall m$, $Q_k = \mu_k = 20$ dBm $\forall k$, $\sigma_k^2 = \eta_m^2 = N_o B$ $\forall k,m$, $N_o = -174$ dBm/Hz, and $B = 20$ MHz. This means $(M/L)P$ will be the power budget of each BS. Besides, we set $N = 5$ for the user-AP association.

\begin{figure}[t]
    \centerline{\includegraphics[scale = 1]{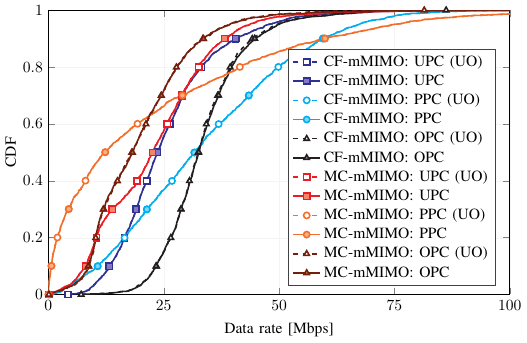}}    
    \caption{DL user rate under (a) CF-mMIMO and (b) MC-mMIMO with UPC, PPC, and OPC.}
    \label{fig:3}   
    \centerline{\includegraphics[scale = 1]{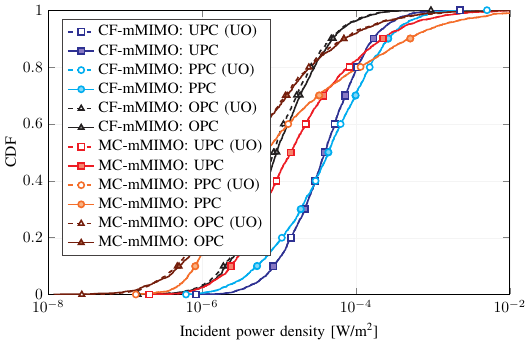}}    
    \caption{IPD user radiation under (a) CF-mMIMO and (b) MC-mMIMO with UPC, PPC, and OPC.}
    \label{fig:4}
    \vspace{-3mm}
\end{figure}

In line with \cite{Chi21}, we focus on whole-body EMF constraints applied to the general public: $I_k = 10$ W/m\textsuperscript{2} $\forall k$ for the IPD limit and $E_{k,n} = 0.08$ W/kg $\forall k,n$ for the SAR (single) metric, with coefficient $b_{k,n} = 8$ kg\textsuperscript{-1} $\forall k,n$.

Finally, we assume a coherence time and bandwidth of $1$ ms and $200$ kHz, respectively \cite{Elw23}. Therefore, $\tau_c = 200$ symbols or (time-frequency samples) are available for communication. Accordingly, the first $\tau_p = K/2$ symbols will be dedicated to channel estimation, i.e., two users will be sharing the same pilots, and $\tau_d = \tau_u = (\tau_c - \tau_p)/2$ samples will be assigned to both DL and UL phases (cf. \eqref{eq:9} and \eqref{eq:11}).

\subsection{Downlink Performance}
Along with the optimal power control (OPC) derived in Subsection~\ref{sec:4.1}, we include uniform power control (UPC) and proportional power control (PPC) as benchmark schemes \cite{Dem21}. This will help to emphasize the performance of our proposal. Recall that although all designs are based on CSI uncertainties, they are ultimately applied to the true channels.

\begin{figure}[t]
    \centerline{\includegraphics[scale = 1]{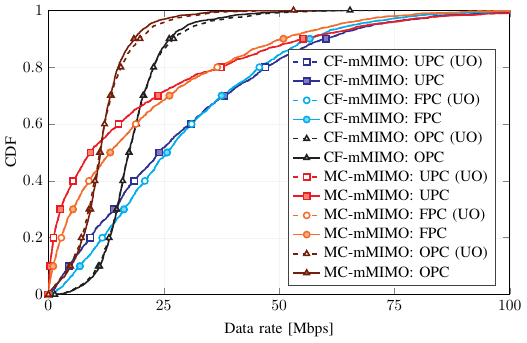}}
    \caption{UL user rate under (a) CF-mMIMO and (b) MC-mMIMO with UPC, FPC, and OPC.}
    \label{fig:5}        
    \centerline{\includegraphics[scale = 1]{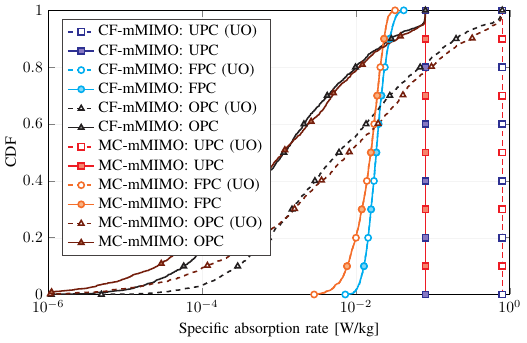}}
    \vspace{0.125em}
    \caption{SAR user radiation under (a) CF-mMIMO and (b) MC-mMIMO with UPC, FPC, and OPC.}
    \label{fig:6}
    \vspace{-3mm}
\end{figure}

The rate CDF for $K = 20$, $L = 4$, and $M = 40$ is depicted in Fig.~\ref{fig:3}, where the EMF-unconstrained optimization (UO) is also shown. As expected, CF-mMIMO outperforms the multi-cell approach in all cases, especially for unlucky users. 

Our power control mechanism also ensures a higher QoS when compared to the other two strategies. This improvement is more notorious in the cell-free deployment: around $50\%$ and $80\%$ of the users w.r.t. PPC and UPC, respectively.

Additionally, notice that for this particular setting, the EMF limitation does not seem detrimental to the QoS (the rate performance remains almost unaltered). This is due to the high propagation losses in the DL, which again stresses the safety of both mMIMO architectures. As we will see later, a smaller threshold should be considered for the radiation constraint to decrease the user rate significantly.

The previous phenomena can be better understood with the CDF of the radiation per user (measured as IPD), presented in Fig.~\ref{fig:4}. We can observe that the exposure barely changes when removing the EMF constraint. Although larger distances in the MC-mMIMO lead to lower values, we are still far from the $10$ W/m\textsuperscript{2} limit, and the OPC generally yields smaller IPDs.

\subsection{Uplink Performance}
The CDFs of the data rate and the perceived EMF (in terms of SAR) for $K = 20$, $L = 4$, and $M = 40$ are depicted in Figs.~\ref{fig:5} and \ref{fig:6}, respectively. Fractional power control (FPC) will be a heuristic baseline instead of PPC \cite{Dem21}. 

\begin{figure}[t]    
    \centerline{\includegraphics[scale = 1]{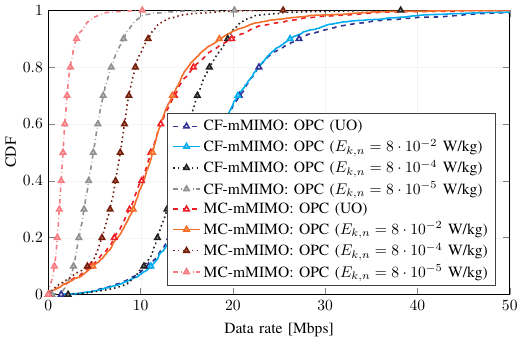}}    
    \caption{UL user rate under (a) CF-mMIMO and (b) MC-mMIMO with OPC for different $E_{k,n}$.}
    \label{fig:7}    
    \centerline{\includegraphics[scale = 1]{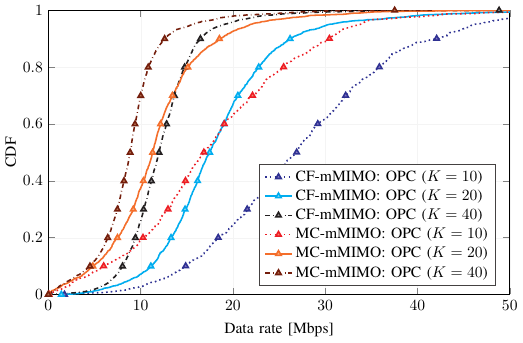}}    
    \caption{UL user rate under (a) CF-mMIMO and (b) MC-mMIMO with OPC for different $K$.}
    \label{fig:8}
\end{figure}

Once more, the cell-free solution provides larger rates than its centralized counterpart, and the OPC outperforms the rest of the power control mechanisms. Besides, the unconstrained optimization yields practically the same QoS. However, unlike before, SAR increases considerably in the UO setup (even exceeding the ultimate $0.08$ W/kg limit), and a similar exposure is perceived in both mMIMO scenarios. This highlights the interest in EMF analysis for the UL. 

To further investigate the role of radiation, the data rate w.r.t. OPC is depicted in Fig.~\ref{fig:7} for different EMF thresholds. As we can see, the performance only worsens when the limit is substantially lesser than the standard established by ICNIRP and FCC (solid line), which indicates we can satisfy the health regulations without compromising the data rate. De novo, CF-mMIMO generates better QoS in all situations.

Finally, to compare the results of both multi-antenna technologies versus the user load, in Fig.~\ref{fig:8}, we show the data rate obtained under OPC for different values of $K$. Unsurprisingly, the QoS deteriorates substantially with a large number of users (e.g., $K = 40$), yet the distributed design always surpasses the multi-cell scheme. As a result, one can infer that cell-free approaches are more suited for handling denser deployments.

\section{Conclusions}\label{sec:6}
This paper addresses the problem of designing power control techniques to boost QoS under EMF constraints in a user-centric CF-mMIMO network. The perceived radiation in the DL and UL scenarios has been modeled through IPD and SAR metrics, respectively. Numerical results demonstrate that our decentralized approach can outperform the MC-mMIMO architecture and that, although the UL is more sensitive to the exposure limitations, their impact is less prominent in the DL.

\bibliographystyle{IEEEtran}
\bibliography{IEEEabrv,References}

\end{document}